\documentclass[a4paper,man]{apa6}

\usepackage[square,numbers]{natbib}
\usepackage{amssymb}
\usepackage{graphicx}
\usepackage{amsmath}
\usepackage{amscd}
\usepackage{mathrsfs}
\usepackage{latexsym}
\usepackage{amsthm}
\usepackage{bbm}
\usepackage{graphicx}
\usepackage{appendix}
\usepackage{float}
\usepackage{subfigure,tikz}
\usepackage{circuitikz}
\usetikzlibrary{intersections}

\usepackage{pdflscape}
\usepackage{longtable}
\usepackage{rotating}

\usepackage{url,lineno}
\usepackage[english]{babel}
\usepackage[utf8x]{inputenc}
\usepackage{amsmath}
\usepackage{graphicx}
\usepackage[colorinlistoftodos]{todonotes}

\title{Towards a Quantum Theory of Humour}
\shorttitle{Quantum Theory of Humour}

\twoauthors{Liane Gabora}{Kirsty Kitto}

\twoaffiliations{Department of Psychology, University of British Columbia \\ Fipke Centre for Innovative Research, 3247 University Way, Kelowna, BC \\ Canada, V1V 1V7 \\
liane.gabora@ubc.ca \\ https://people.ok.ubc.ca/lgabora/}{Science and Engineering Faculty, Mathematical Sciences \\ Queensland University of Technology, Brisbane, Australia \\ kirsty.kitto@qut.edu.au \\ http://staff.qut.edu.au/staff/kittok/\\ \bigskip
Reference: \\Gabora, L. \& Kitto, K. (2016). Towards a quantum theory of humour. Frontiers in Physics (section: Interdisciplinary Physics), 4(53). \\ doi: 10.3389/fphy.2016.00053
}

\abstract{This paper proposes that cognitive humour can be modelled using the mathematical framework of quantum theory, suggesting that a Quantum Theory of Humour (QTH) is a viable approach. We begin with brief overviews of both research on humour, and the generalized quantum framework. We show how the bisociation of incongruous frames or word meanings in jokes can be modelled as a linear superposition of a set of basis states, or possible interpretations, in a complex Hilbert space. The choice of possible interpretations depends on the context provided by the set-up versus the punchline of a joke. We apply QTH first to a verbal pun, and then consider how this might be extended to  frame blending in cartoons.  An initial study of 85 participant responses to 35 jokes (and a number of variants) suggests that there is reason to believe that a quantum approach to the modelling of cognitive humour is a viable new avenue of research for the field of quantum cognition.}

\begin{document}
\maketitle

\section{Introduction}\label{intro} 
Humour has been called the ``killer app'' of language \citep{brone_1._2015}; it showcases the speed, playfulness, and flexibility of human cognition, and can instantaneously put people in a positive mood. For over a hundred years scholars have attempted to make sense of the seemingly nonsensical cognitive processes that underlie humour. Despite considerable progress with respect to categorizing different forms of humour (e.g., such as irony, jokes, cartoons, and slapstick) and understanding what people find funny, there has been little investigation of the question: What kind of formal theory do we need to model the cognitive representation of a joke at the instant it is understood? 

This paper attempts to answer this question with a new model of humour that uses a generalization of the quantum formalism. The last two decades have witnessed an explosion of applications of these formalisms to psychological phenomena that feature ambiguity and/or contextuality \citep{khrennikov_ubiquitous_2010,busemeyer_quantum_2012,asano_quantum_2015}.  
Many different psychological phenomena have been studied, including  the combination of words and concepts \citep{aerts_quantum_2009,aerts_theory_2005, aerts_theory_2005-1, gabora_contextualizing_2002, bruza_is_2009, bruza_probabilistic_2015}, similarity and memory \citep{pothos_quantum_2013, nelson_how_2013}, information retrieval \citep{van_rijsbergen_geometry_2004, melucci_basis_2008}, decision making and probability judgement errors \citep{aerts_applications_1997, busemeyer_quantum_2006, busemeyer_quantum_2011-1, lambert_mogiliansky_type_2009, yukalov_processing_2009}, vision \citep{atmanspacher_quantum_2004, atmanspacher_necker-zeno_2013}, sensation--perception \citep{khrennikov_quantum-like_2015}, social science \citep{haven_quantum_2013, kitto_attitudes_2013}, cultural evolution \citep{gabora_cognitive_2001, gabora_model_2009-1}, and creativity \citep{gabora_study_2015, swan_concept_2013}. 


These quantum inspired approaches make no assumption that phenomena at the quantum level affect the brain; they draw solely on abstract formal structures that, as it happens, found their first application in quantum mechanics. The common approach is to utilize the structurally different nature of quantum probability. While in classical probability theory events are drawn from a common sample space, quantum models define states and variables with reference to a context, represented using a basis in a Hilbert space. This results in behaviour such as interference, superposition and entanglement, and ambiguity with respect to the outcome is resolved with a quantum measurement and a collapse to a definite state. 

This makes the quantum inspired approach an interesting new candidate for a theory of humour. Humour often involves ambiguity due to the presence of incongruous schemas: internally coherent but mutually incompatible ways of interpreting or understanding a statement or situation.  As a simple example, consider the following pun: 
\begin{quotation}
``Time flies like an arrow. Fruit flies like a banana.''
\end{quotation} 
This joke hangs on the ambiguity of the phrase FRUIT FLIES, where the word FLIES can be either a verb or a noun. As a verb, FLIES means ``to travel through the air''. However, as a noun, FRUIT FLIES are ``insects that eat fruit''. Quantum formalisms are highly useful for describing cognitive states that entail this form of ambiguity. This paper will propose that the quantum approach enables us to naturally represent the process of ``getting a joke''.

We start by providing  a brief overview of the relevant research on humour.

\section{Brief Background in Humour Research}
\label{sec:humour}

Even within psychology, humour is approached from multiple directions. Social psychologists investigate the role of humour in establishing, maintaining, and disrupting social cohesion and social status, developmental psychologists investigate how the ability to understand and generate humour changes over a lifetime, and health psychologists investigate possible therapeutic aspects of humour. This paper deals solely with the cognitive aspect of humour. 
Much cognitive theorizing about humour assumes that it is driven by the simultaneous perception \citep{attardo_linguistic_1994,raskin_semantic_2012} or `bisociation' \citep{koestler_act_1964} of incongruent \emph{schemas}. Schemas can be either static \emph{frames}, as in a cartoon, or dynamically unfolding \emph{scripts}, as in a pun. For example, in the ``time flies'' joke above, interpreting the phrase FRUIT FLIES as referring to the insect is incompatible with interpreting it as food travelling through the air. Incongruity is generally accompanied by the violation of expectations and feelings of surprise. While earlier approaches posited that humour comprehension involves the \emph{resolution} of incongruous frames or scripts \citep{shultz_order_1974,suls_two-stage_1972}, the notion of resolution often plays a minor role in contemporary theories, which tend to view the punchline as activating multiple schemas simultaneously and thereby underscoring ambiguity (e.g., \citet{martin_psychology_2010}; \citet{mcgraw_benign_2010}).

There are computational models of humour detection and understanding (e.g., \citet{reyes_multidimensional_2013}), in which the interpretation of an ambiguous word or phrase generally changes as new surrounding contextual information is parsed. For example, in the ``time flies'' joke, this kind of model would shift from interpreting FLIES as a verb to interpreting it as a noun. There are also computational models of humour that  generate jokes through lexical replacement; for example, by replacing a `taboo' word with a similar-sounding innocent word (e.g.,\citep{binsted_childrens_1997};\citet{valitutti__2013}). 
These computational approaches to humour are interesting, and  occasionally generate jokes that are laugh-worthy. However, while they tell us something about humour, we claim that they do not provide an accurate model of the cognitive state of a human mind at the instant of perceiving a joke. As mentioned above, humour psychologists believe that humour often involves not just shifting from one interpretation of an ambiguous stimulus to another, but simultaneously holding in mind the interpretation that was perceived to be relevant during the set-up and the interpretation that is perceived to be relevant during the punchline. For this reason, we turned to the generalized quantum formalism as a possible approach for modelling the cognitive state of holding two schemas in mind simultaneously.  

\section{Brief Background in Generalized Quantum Modeling}\label{sec:quantum}

Classical probability describes events by considering subsets of a common sample space \citep{isham_lectures_1995}. That is, considering a set of elementary events, we find that some event $e$ occurred with probability $p_e$. Classical probability arises due to a lack of knowledge on the part of the modeller. The act of measurement merely reveals an existing state of affairs; it does not interfere with the results. 

In contrast, quantum models use variables and spaces that are defined with respect to a particular context (although this is often done implicitly). Thus, in specifying that an electron has spin `up' or `down', we are referring to experimental scenarios (e.g., Stern-Gerlach arrangements and polarizers) that denote the context in which a measurement occurred. This is an important subtlety, as many experiments have shown that it is impossible to attribute a pre-existing reality to the state that is measured; measurement necessarily involves an interaction between a state and the context in which it is measured, and this is traditionally modelled in quantum theory using the notion of projection.  The \emph{state} \(| \Psi \rangle\) representing some aspect of interest in our system is written as a linear superposition of a set of \emph{basis states} \(\{| \phi_i \rangle\}\) in a \emph{Hilbert space} \(\mathcal{H}\) which allows us to define notions such as distance and inner product. In creating this superposition we weight each basis state with an amplitude term, denoted \(a_i\), which is a complex number representing the contribution of a component basis state \(|\phi_i\rangle\) to the state \(|\Psi\rangle\). Hence \(| \Psi \rangle = \sum_ia_i | \phi_i \rangle\). The square of the absolute value of the amplitude equals the probability that the state changes to that particular basis state upon measurement. This non-unitary change of state is called \emph{collapse}. The choice of basis states is determined by the \emph{observable}, \(\hat{O} \), to be measured, and its possible outcomes \(o_i\). The basis states corresponding to an observable are referred to as \emph{eigenstates}. Observables are represented by self-adjoint operators on the Hilbert space. 
Upon \emph{measurement}, the state of the entity is projected onto one of the eigenstates. 

It is also possible to describe combinations of two entities within this framework, and to learn about how they might influence one another, or not.
Consider two entities $A$ and $B$ with Hilbert spaces $\mathcal{H_A}$ and $\mathcal{H_B}$. We may define a basis  ${| i \rangle_A}$ for $\mathcal{H_A}$ and a basis  ${| j \rangle_B}$ for $\mathcal{H_B}$, and denote the amplitudes associated with the first as \(a_i^A\) and the amplitudes associated with the second as \(a_j^B\). 
The Hilbert space in which a composite of these entities exists is given by the tensor product  $\mathcal{H_A} \otimes \mathcal{H_B}$.
  The most general state in $\mathcal{H_A} \otimes \mathcal{H_B}$ has the form 
\begin{equation} 
| \Psi \rangle_{AB} = {\sum}_{i,j} a_{ij} | i \rangle_A \otimes | j \rangle_B\
\end{equation}
This state is separable if $a_{ij}=a_i^Aa_j^B$.
It is inseparable, and therefore an entangled state, if  $a_{ij}\neq a_i^Aa_j^B$.

In some applications the procedure for describing entanglement is more complicated than what is described here. For example, it has been argued that the quantum field theory procedure, which uses Fock space to describe multiple entities, gives a kind of internal structure that is superior to the tensor product for modelling concept combination \citep{aerts_quantum_2009}. Fock space is the direct sum of tensor products of Hilbert spaces, so it is also a Hilbert space. For simplicity, this initial application of the quantum formalsm to modelling humour will omit such refinements, but such a move may become necessary in further developments of the model. 

Quantum models can be useful for describing situations involving \emph{potentiality}, in which change of state is nondeterministic and contextual. The concept of potentiality has broad implications across the sciences; for example, every biological trait not only has direct implications for existing phenotypic properties such as fitness, but both enables and constrains potential future evolutionary changes for a given species. 
The quantum approach been used to model the biological phenomenon of \emph{exaptation} --- wherein a trait that originally evolved for one purpose is co-opted for another (possibly after some modification) \citep{gabora_quantum_2013}. The term \emph{exaptation} was coined by \citet{gould_exaptationmissing_1982} to denote what Darwin referred to as \emph{preadaptation}.\footnote{The terms \emph{exaptation}, \emph{preadaptation} and \emph{co-option} are often used interchangeably.} 
Exaptation occurs when selective pressure causes this potentiality to be exploited. Like other kinds of evolutionary change, exaptation is observed across all levels of biological organization, i.e., at the level of genes, tissue, organs, limbs, and behavior. 
Quantum models have also been used to model the cultural analog of exaptation, wherein an idea that was originally developed to solve one problem is applied to a different problem \citep{gabora_quantum_2013}. For example, consider the invention of the tire swing. It came into existence when someone re-conceived of a tire as an object that could form the part of a swing that one sits on. This re-purposing of an object designed for one use for use in another context is referred to as \emph{cultural exaptation}. Much as the current structural and material properties of an organ or appendage constrain possible re-uses of it, the current structural and material properties of a cultural artefact (or language, or art form, etc.) constrain possible re-uses of it. Here, we suggest that incongruity humour constitutes another form of exaptation; an ambiguous word, phrase, or situation, that was initially interpreted one way is revealed to have a second, incongruous interpretation.

\section{A Quantum Inspired Model of Humour}\label{sec:quantumModel}

A quantum theory of humour (QTH) could potentially inherit several core of from previous cognitive theories of humour while providing a unified underlying model. Considering the past work discussed in section~\ref{sec:humour}, it seems reasonable to focus on the notion that cognitive humour involves an ambiguity brought on by the bisociation of internally consistent but mutually incongruous schemas. Thus,  cognitive humour appears to arise from the double think that is brought about by being forced to reconsider some currently held interpretation of a joke within the light of new information: a frame shift. 
Such an insight opens up humour to quantum-like models, as a frame shift of an ambiguous concept is well modelled by the notion of a quantum superposition described using two sets of incompatible basis states within some underlying Hilbert space structure.

In what follows we shall introduce some features that would be required in a formal QTH as we start to sketch out a preliminary quantum inspired model of humour.  This initial model provided enough insight to formulate  an experimental procedure aimed at discovering whether humour was likely to behave in a quantum-like manner. A  preliminary study is introduced in section~\ref{sec:experiment}, which has provided insights and opened up a rich set of avenues for future investigation. Thus, this paper is a first step towards the development of a formal theory for a field that to date has not been well modelled. 

\subsection{The Mathematical Structure of QTH}
We start our journey toward a QTH by building upon an existing model of conceptual combination first proposed by \citet{gabora_contextualizing_2002}: the State--COntext--Property (SCOP) model. As per the standard approach used in most quantum-like models of cognition, \(|\Psi \rangle\)  represents the state of an ambiguous  element, be it a word, phrase, object, or something else, and its different possible interpretations are represented by basis states. Core to the SCOP model is a treatment of  the context in which every measurement of a state occurred, and the resultant property that was measured. These three variables are stored as a triple in a lattice. 

\subsubsection{The State Space.}
Following \citet{aerts_theory_2005}, the set of all possible interpretation states for the ambiguous element of a joke is given by a state space $\Sigma$. Specific interpretations of a joke are denoted by $|p\rangle, |q\rangle, |r\rangle, \dots \in \Sigma$ which form a basis in a complex  Hilbert space \(\mathcal{H}\).
Before the ambiguous element of the joke is resolved, it is in a state of potentiality, represented by a superposition state of all possible interpretations. Each of these represents a possible understanding arising due to activation of a schema associated with a particular interpretation of an ambiguous word or situation. The interpretations that are most likely are most heavily weighted.
The amplitude term associated with each basis state represented by a complex number coefficient \(a_i\) gives a measure of how likely an interpretation is given the current contextual information available to the listener.  
We assume that all basis states have unit length, are mutually orthogonal, and are complete, thus \(\sum_i|a_i|^2 = 1\).


\subsubsection{The Context.} In the context of a traditional verbal joke the context consists primarily of the setup, and the setup is the only contextual element considered in the study in Section~\ref{sec:quantumModel}. However, it should be kept in mind that several other contextual factors not considered in our analysis can affect perceived funniness. Prominent amongst these is the delivery; the way in which a joke is delivered can be everything when it comes to whether or not it is deemed funny. Other factors include the surroundings, the person delivering the joke, the power relationships among different members of the audience, and so forth. 

As a first step, we might represent the set of possible contexts for a given joke as \(c_i \in \mathcal{C}\). Each possible interpretation of a joke comes with a set \(f_i \in \mathcal{F}\) of features (or properties), which may be weighted according to their relevance with respect to this contextual information.  The \emph{weight} (or renormalized applicability) of a certain property given a specific interpretation $| p \rangle$ in a specific context $c_i \in \mathcal{C}$ is given by $\nu$. 
For example, $\nu (p, f_1)$ is the weight of feature ${f_i}$ for state $| p \rangle$, which is determined by a function from the set 
$ \Sigma  \times {\mathcal F} $ to the interval $[0, 1]$. 
We write:
\begin{align}\
\nu: \Sigma  \times {\mathcal F} &\rightarrow [0, 1] \\ 
(p, f_i) &\mapsto \nu(p, f_i).
\nonumber
\end{align}

\subsubsection{Transition probabilities.}
A second function $\mu$ describes the transition probability from one state to another under the influence of a particular context. For example, $\mu$\textit{(q, e, p)} is the probability that state $|p\rangle$ under the influence of context $c_i$ changes to state $|q\rangle$. Mathematically, $\mu$ is a function from the set $\Sigma \times \mathcal{C} \times \Sigma $
to the interval $[0, 1]$, where $\mu(q, e, p)$ is the probability that state $|p\rangle$ under the influence of context $|e\rangle$ changes to state $|q\rangle$. We write: 
\begin{align}
\mu: \Sigma \times \mathcal{C} \times \Sigma 
&\rightarrow [0, 1]  \\
(q, e, p) &\mapsto \mu(q, e, p). 
\nonumber
\end{align}
Thus, a first step towards a full quantum model of humour consists of the 3-tuple $(\Sigma, \mathcal{C}, \mathcal{F})$, and the functions $\nu$ and $\mu$. 
However, we have yet to address the core questions that should be asked of any cognitive theory of humour: 
what is the underlying cognitive model of the funniness of a joke?

\subsection{The Humour of a Joke}


As the listener hears a joke, more context is provided, and in our model their understanding (i.e., the cognitive state of the listener) evolves according to the transition probabilities associated with the cognitive state and the emerging context. When the listener interprets the joke the listener is perceiving a bisociation of meaning. That is, the first interpretation that the listener ascribes to the joke changes because two meanings are possible for the core concept in the joke. 
A projective measurement onto a funniness frame is the mechanism that we use to model the likelihood that a given joke is considered funny. 

Thus, in our model, funniness plays the role of a measurement operator, and it is affected by the shift that occurs in the understanding of a joke with respect to two possible framings: one created by the setup, and one by the punchline.  The probability of a joke being regarded as funny or not is proportional to the projection of the individual's understanding of the joke ($|\Psi\rangle$) onto a basis representing funniness. This means that the probability of a joke being considered as funny, $p_F$ is given by a projection onto the $|1\rangle$ axis in \(\mathcal{H}_F^2\), a 2 dimensional Hilbert sub-space where $|0\rangle$ represents `not funny' and $|1\rangle$ represents `funny'. 
\begin{equation}
p_{F}=| |1\rangle\langle 1| \Psi\rangle |^2
\end{equation}
Similarly, the probability of a joke being regarded as not funny is represented by 
\begin{equation}
p_{\bar{F}}=| |0\rangle\langle 0| \Psi\rangle |^2.
\end{equation}
Note that $|\Psi\rangle$ evolves as the initial conceptualisation of the joke is reinterpreted with respect to the frame of the punchline. This is a difficult process to model, and we consider the work in this paper to be an early first step towards an eventually more comprehensive theory of humour that includes predictive models. 

To start in this journey towards a QTH, we will now present two examples in which two specific instances of humour are considered within the perspective of this basic quantum inspired model. First the approach is applied to a pun. Second, the approach is applied to a cartoon that is a frame blend. Both scenarios will help to deepen our understanding of the significant complexity of humour, and the difficulties associated with creating a mathematical model of this important human phenomenon. 

\subsection{Example 1: A Pun}
\label{sec:pun}

Consider the pun: ``Why was 6 afraid of 7? Because 789.'' The humour of this pun hinges on the fact that the pronunciation of the number EIGHT, a noun, is identical to that of the verb ATE. We refer to this ambiguous word, with its two possible meanings, as EYT.
An individual's interpretation of the word EYT is represented by \(|\Psi\rangle\), a vector of length equal to 1. This is a linear superposition of basis states in the semantic sub-space \(\mathcal{H}^2_M\) which represents possible states (meanings) of the word EYT: EIGHT or ATE.\footnote{We acknowledge that other interpretations are possible, and so this is a simplified model. It is straightforward to extend the model into higher dimensions by adding further interpretations as basis states.}  
The interpretation of EYT as a noun, and specifically the number EIGHT, will be denoted by the unit vector \(|n\rangle\). The verb interpretation, ATE, is denoted by the unit vector \(|v\rangle\). The set $\{|n\rangle,|v\rangle\}$ forms a basis in \(\mathcal{H}^2_M\).  Thus, we have now expanded our original 2-dimensional funniness space with an additional 2-dimensional semantic space, where the full space $\mathcal{H}^4=\mathcal{H}^2_F\otimes \mathcal{H}^2_M$. We note that these two spaces should not be considered as mutually orthogonal, but that they will overlap. If they were orthogonal then the funniness of a joke would be independent of the interpretation that a person attributes to it. 

\begin{figure}[h!]
\centering
    \begin{tikzpicture}
		[scale=1,line cap=round,
		axes/.style=,
		important line/.style={very thick},
		information text/.style={rounded corners,fill=red!10,inner sep=1ex},
		dot/.style={circle,inner sep=1pt,fill,label={#1},name=#1}			
		]	
		\draw[black,->,ultra thick] (0, 0) -- node[pos=1.2] {$|\Psi\rangle$} (1.5, 3);    
        \draw[black,->,thick] (0,0) -- node[pos=1.2]{$|n\rangle$}(0,3.5);
        \draw[black,->,thick] (0,0) -- node[pos=1.2]{$|v\rangle$}(3.5,0);
        \draw[dotted,black,->,thick] (1,1.6) to[bend left] (1.8,0.5);
   \coordinate (FIRST NE) at (current bounding box.north east);
   \coordinate (FIRST SW) at (current bounding box.south west);
	\end{tikzpicture}\hspace*{1cm}
    \begin{tikzpicture}
		[scale=1,line cap=round,
		axes/.style=,
		important line/.style={very thick},
		information text/.style={rounded corners,fill=red!10,inner sep=1ex},
		dot/.style={circle,inner sep=1pt,fill,label={#1},name=#1}	  	
		]		   
        
       \useasboundingbox (FIRST SW) rectangle (FIRST NE);
		\draw[black,->,ultra thick] (0, 0) -- node[pos=1.2] {$|\Psi\rangle$} (3, 0.5);    
        \draw[black,->,thick] (0,0) -- node[pos=1.2]{$|0\rangle$}(1,3.5);
        \draw[black,->,thick] (0,0) -- node[pos=1.2]{$|1\rangle$}(3.5,-1);
        \draw[dotted,black,->,thick] (3,0.5) to (2.65,-0.6);
	\end{tikzpicture}
    \hspace*{1cm}
    \begin{tikzpicture}
		[scale=1,line cap=round,
		axes/.style=,
		important line/.style={very thick},
		information text/.style={rounded corners,fill=red!10,inner sep=1ex},
		dot/.style={circle,inner sep=1pt,fill,label={#1},name=#1}	  	
		]		   
        
       \useasboundingbox (FIRST SW) rectangle (FIRST NE);
		\draw[black,->,ultra thick] (0, 0) -- node[above, pos=1.2] {$|\Psi\rangle$} (3, -0.85);    
        \draw[black,->,thick] (0,0) -- node[pos=1.2]{$|0\rangle$}(1,3.5);
        \draw[black,->,thick] (0,0) -- node[pos=1.2]{$|1\rangle$}(3.5,-1);
        \draw[black,->,thick] (0,0) -- node[pos=1.2]{$|n\rangle$}(0,3.5);
        \draw[black,->,thick] (0,0) -- node[pos=1.2]{$|v\rangle$}(3.5,0);
	\end{tikzpicture}\\\vspace*{1cm}
    (a)\hspace*{6cm}(b)\hspace*{6cm}(c)
    \caption{The humour of a joke can be explained as arising from a 
measurement process that occurs with respect to two incompatible frames. Using the example of the pun, 
(a) 
the meaning of the set-up is reinterpreted with EYT updating 
towards the interpretations ATE. 
(b) Funniness is then treated as a measurement, with the probability of funniness being judged with respect to a projection on the $\{|0\rangle,|1\rangle\}$ basis. In this case there is a large probability of the joke being considered funny due to the dominant component of the projection of $|\Psi\rangle$ lying on the $|1\rangle$ axis.
(c) The cognitive state of the subject then collapses to the observed state (i.e. funny or not). 
 }\label{fig:humour}
\end{figure}
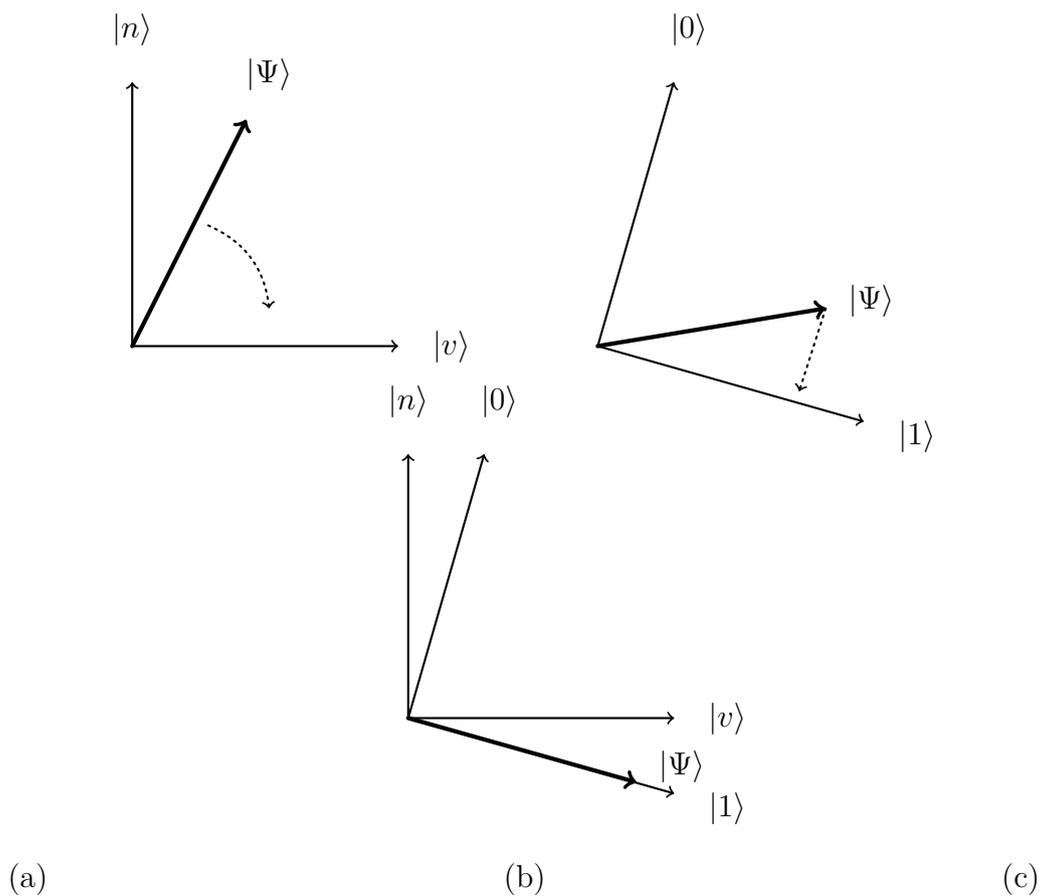

With this added mathematical structure, we can represent the interpretation of the joke as a superposition state in \(\mathcal{H}^2_M\)
\begin{equation} \label{eq:punState}
|\Psi\rangle = a_n|n\rangle + a_v|v\rangle,
\end{equation}
where \(a_n\) and \(a_v\) are amplitudes which, when squared, represent the probability of a listener interpreting the joke in a noun or a verb form (\(|n\rangle\) and \(|v\rangle\)) respectively. This state is depicted in Fig.~\ref{fig:humour}(a), which shows a superposition state in the semantic space. When given no context in the form of the actual presentation of the joke, these amplitudes represent the prior likelihood of a listener interpreting the uncontextualized word (i.e. EYT) in either of the noun or verb senses (e.g. a free association probability, see \cite{nelson_how_2013} for a review). However, we would expect to see these probabilities evolving throughout the course of the pun as more and more context is provided (in the form of additional sentence structure). Throughout the course of the joke, the state vector $|\Psi\rangle$ therefore evolves to a new position in $\mathcal{H}^4$.

Since the set-up of the joke,``Why was 6 afraid of 7?'', contains two numbers, it is likely that the \underline{numbers} interpretation of EYT is activated (a situation represented in Figure~\ref{fig:humour}(a)). The listener is biased towards an interpretation of EYT in this sense, and so we would expect that $a_n>>a_v$. 
However, a careful listener will feel confused upon considering this set-up because we do not think of numbers as beings that experience fear. This keeps the interpretation of EYT shifted away from an equivalence with the eigenvector $|n\rangle$.  
As the joke unfolds, the \underline{predator} interpretation that was hinted at in the set-up by the word ``afraid'', and reinforced by ``789'', activates a more definite alternative meaning, ATE represented by  $|v\rangle$. This generates an alternative  interpretation of the punchline: that the number 7 ate the number 9. The cognitive state $|\Psi\rangle$ has evolved to a new position in $\mathcal{H}^4$, a scenario that is represented in Figure~\ref{fig:humour}(b). At this point a measurement occurs: the individual 
either considers the joke 
as funny or not within the context represented by the funniness sub space $\mathcal{H}^2_F$, and a collapse to the relevant funniness basis state occurs (see Figure~\ref{fig:humour}(c)). Note that this final state still contains a superposition within the meaning subspace $H_M^2$ --- the funniness judgement merely shifts the interpretation of the joke, it does not eliminate the bisociation. Rather, it depends upon it.
 

If we consider the set of properties associated with EYT then we would expect to see two very different prototypical characteristics associated with each interpretation. For example, the EIGHT interpretation is difficult to map into properties such as `food' denoted $f_1$, and  `not living' denoted $f_2$ (since when something is eaten it is usually no longer alive). Because `food' and `not living'  are not properties of EIGHT, $\nu(p, f_0) << \nu(n, f_0)$, and similarly $\nu(p, f_1) << \nu(n, f_1)$.
However, `food' and `not living' are properties of EYT, $\nu(p, f_0) << \nu(v, f_0)$, and similarly $\nu(p, f_1) << \nu(v, f_1)$. 

We can now start to construct a model of humour that could be correlated with  data. If jokes satisfy the law of total probability (LTP) then their funniness should satisfy the distributive axiom, which states that the total probability of some observable should be equal to the sum of the probabilities of it under a set of more specific conditions. Thus, considering a funniness observable $\hat{O}_F$ (with eigenstates $\{|1\rangle,|0\rangle\}$ and the semantic observable $\hat{O}_M$ (with a simplified two eigenstate structure $\{|M\rangle,|\bar{M}\rangle\}$ representing two possible meanings that could be attributed to the joke). We can take the spectral decomposition of $\hat{O}_M = m| M\rangle\langle M| + \bar{m}| \bar{M}\rangle\langle \bar{M}|$, where $m,\bar{m}$ are eigenvalues of the two eigenstates $\{|M\rangle,|\bar{M}\rangle\}$. Doing this, we should find that if this system satisfies the LTP then the probability of the joke being judged as funny is equal to the sum of the probability of it being judged funny 
{\emph given either semantic interpretation}
\begin{equation}\label{eq:LTP}
p(F)= p(|1\rangle)=p(M)\cdot p(F|M) + p(\bar{M})\cdot p(F|\bar{M}).
\end{equation}
We can manipulate the interpretation that a participant is likely to attribute to a joke by changing the semantics of the joke itself. Thus, changing the joke should change the semantics and so affect the humour that is attributed to the joke. 
We shall return to this idea in Section~\ref{sec:experimentsHumour}.

This section has demonstrated that a formal approach to concept interactions which has been previously shown to be consistent with human data \citep{aerts_quantum_2009} can be adapted to the simultaneous perception of incongruous meanings of an ambiguous word or phrase in the understanding of a pun. What other aspects of cognitive humour might this new mathematical apparatus be applied to?

\subsection{Example 2: A Frame Blend}

\begin{figure} \label{fig:fig-cartoon}
\begin{center}
 	\includegraphics[height=8cm]{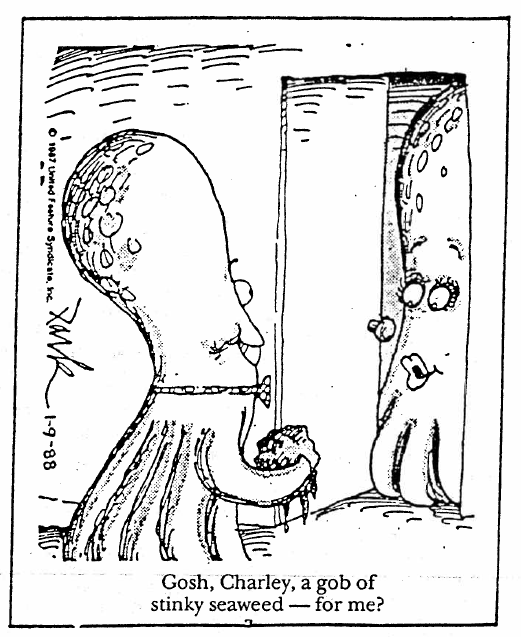}\includegraphics[height=8cm]{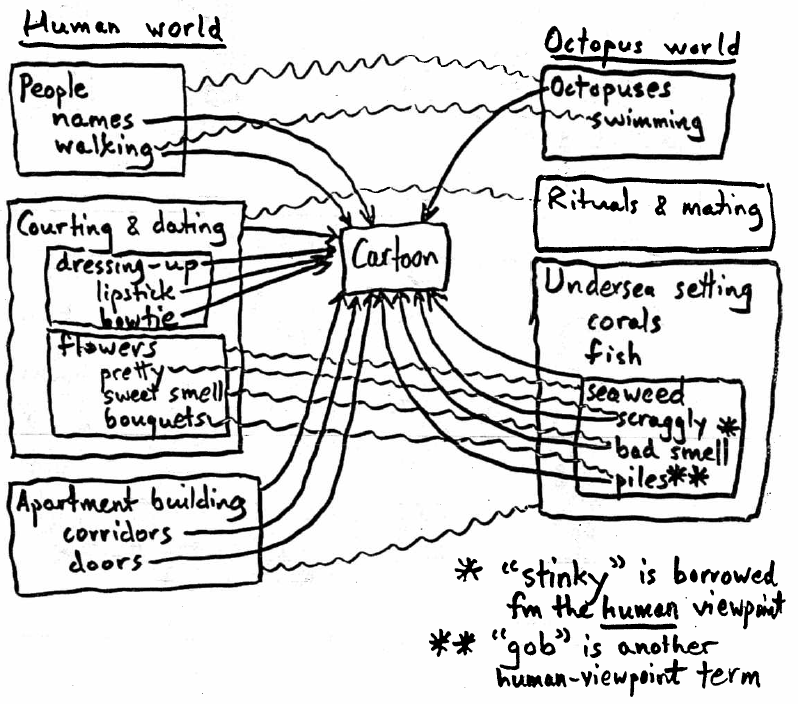}\\
    \hspace{1cm}(a) \hspace{7 cm}(b)\hspace*{2.5cm}
\end{center}
\caption
{ An example of a Frame blend in the QTH. 
(a) An \textit{Off the Leash} cartoon by W. B. Park that blends two frames: a human courtship frame and an octopus frame.  (b) An illustration of how the human courtship frame and the octopus frame map onto one another in the frame blend. (Both examples first appeared in \cite{hofstadter_synopsis_1989}).
}\label{fig:fig-frame-blend}
\end{figure}

Although our first example used a pun for simplicity, we believe that quantum inspired models can be fruitfully applied to more elaborate forms of humour, such as jokes involving incompatible frames or scripts. Our second example is a cartoon that was initially analysed in terms of the concept of a frame blend, which involves the merging of incongruous frames \citep{hofstadter_synopsis_1989}. The cartoon is shown in Figure~\ref{fig:fig-frame-blend}(a) and the frame blend analysis is shown in Figure~\ref{fig:fig-frame-blend}(b). 

In a QTH, the two interpretations of the incongruous situation represented by the scene in Figure~\ref{fig:fig-frame-blend}, as a dating scene and as an octopus scene, would be designated by the unit vectors \(\{|d\rangle,|o\rangle\)\}. 
The cognitive state of perceiving the blended frames is represented as a superposition of the two frames, however the underlying dynamics behind this joke are likely to be different.
That is, rather than being led ``down the garden path'' by the setup and subsequent re-interpretation in light of the punchline, in this scenario the humour appears to result from the immediate simultaneous presentation of seemingly incompatible frames, which creates a similar tension or bisociation as eventually arose in the previous example. As with phenomena such as conceptual combination, in this scenario there are likely to be constraints on how the frames can be successfully blended, and it will be necessary to consider this when constructing a model of humour for this scenario.  
We reserve further exploration of this interesting class of humour for future work. 

\section{Probing the State Space of Humour}
\label{sec:experimentsHumour}

Returning to the question raised by equation \eqref{eq:LTP}, a QTH should be justified by considering whether humour does indeed violate the Law of Total Probability (LTP) \citep{busemeyer_quantum_2012}. However, the complexity of language makes it difficult to test how humour might violate the LTP using a method similar to those followed for decision making  \citep{pothos_quantum_2013}. 
The model discussed above brings us to a position where past work on humour is unlikely to yield the data that is required to perform tests such as this. For example, we currently have no experimental understanding of how the semantics of a joke interplays with its perceived funniness. It seems reasonable to suppose that the two are related, but how? We are not aware of any data sets that provide a way in which to evaluate this relationship. This is problematic, as there are a number of interdependencies in the framing of a joke that make it difficult to construct a model (even before considering factors such as the context in which the joke is made, and the socio-cultural background of the teller and the listener).
In this section we present results from an exploratory study designed to start unpacking whether humour should indeed be considered within the framework of quantum cognition. As an illustrative example, consider the following joke:
\begin{quotation}
$V_O$: ``Time flies like an arrow. Fruit flies like a banana.''
\end{quotation} 
As with the joke discussed in section~\ref{sec:pun}, the humour arises from the ambiguity of the words FRUIT and FLIES. The first frame ($F1$, the set-up), leads one to interpret FLIES as a verb and LIKE as a preposition, but the second frame ($F2$, the punchline), leads one to interpret FRUIT FLIES as a noun and LIKE as a verb. A QTH must be able to explain how the funniness of the joke depends upon a shift in the semantic understanding of the two frames, $F1$ and $F2$. 

We now outline a preliminary study that has helped us to explore the state space of humour. 

\subsection{Stimuli}
We collected a set of 35 jokes and for each joke we developed a set of joke variants. A $V_S$ variant consisted of the set-up only for the original ($V_O$ joke). Thus the $V_S$ variant of the $V_O$ joke is 
\begin{quotation}
$V_S$: ``Time flies like an arrow.''
\end{quotation}
A $V_P$ variant consists of the original punchline only. Thus the $V_P$ variant of the $V_O$ joke is 
\begin{quotation}
$V_P$: ``Fruit flies like a banana.''
\end{quotation}

We then considered the notion of a  congruent punchline as one that does not introduce a new interpretation or context for an ambiguous element of the set-up (or punchline). Congruence was achieved by modifying the set-up to make it congruent with the punchline, or by modifying the punchline to make it congruent with the  set-up. Thus, if the set-up makes use of a noun then a congruent modification would still do this (and similarly for the punchline). 

A $V_{CP}$ variant consists of the original set-up followed a congruent version of the punchline. Thus a $V_{CP}$ variant of the $V_O$ joke is: 
\begin{quotation}
$V_{CP}$: ``Time flies like an arrow; time flies like a bird.'' 
\end{quotation}
A $V_{CS}$ variant consists of the original punchline preceded by a congruent version of the set-up. Thus a $V_{CS}$ variant of the $V_O$' joke is
\begin{quotation} 
$V_{CS}$: ``Horses like carrots; fruit flies like a banana.'' 
\end{quotation}
For some jokes we had a fifth kind of variant. A $V_{IS}$ variant consists of the original set-up followed an incongruent version of the punchline that we believed was comparable in funniness to the original. Thus considering the joke discussed in section~\ref{sec:pun}:
\begin{quotation}
$V_{O}$: ``Why was 6 afraid of 7? Because 789.'' 
\end{quotation}
A $V_{IS}$ variant of this joke is:
\begin{quotation}
$V_{IS}$: ``Why was 6 afraid of 7? Because 7 was a six offender.'' 
\end{quotation}

Thus the stimuli consisted of a  questionnaire containing original jokes, and the above variants presented in randomized order.  
The complete collection of jokes and their variants is presented in the Appendix. 

\subsection{Participants}
The participants in this study were 85 first year undergraduate students enrolled in an introductory psychology course at the University of British Columbia (Okanagan campus). They received partial course credit for their participation.

\subsection{Procedure}
\label{sec:experiment}
Participants signed up for the study using the SONA recruitment system, and subsequently responded at their convenience to an online questionnaire hosted by FluidSurveys.
They were informed that the study was completely voluntary, and that they were free to withdraw from the study at any point in time.
They were also informed that the researcher would not have any knowledge of who participated in the study, and that their participation would not affect their standing in the psychology class or relationship with the university. 
Participants were told that the purpose of the study was to investigate humour, and to help contribute to a better understanding the cognitive process of `getting' a joke.
Participants were asked to fill out consent forms.
If they agreed to participate, they were provided a questionnaire consisting of a series of jokes and joke variants (as described above) and asked to rate the funniness of each using a Likert scale, from 1 (not funny) to 5 (hilarious).
The questionnaire took approximately 25 minutes to complete.
They received partial course credit for their participation.  
 
\subsection{Results}
The mean funniness ratings across all participants for the entire collection of jokes and their variants (as well as the jokes and variants themselves) is provided in the Appendix. Table~\ref{tab:variants} provides a summary of this information (the mean funniness rating of each kind of joke variant across all participants) aggregated across all joke sets.  
As expected, the original joke ($O$) was funniest (mean funniness = 2.70), followed by those jokes that had been intentionally modified to be funny: Incongruent Setup ($IS$) (mean funniness = 2.37) and Incongruent Punchline ($IP$) (mean funniness = 2.12). Next in funniness were the jokes that had been modified to eradicate the incongruency and thus the source of the humour: Congruent Setup ($CS$) (mean funniness = 1.41) and Congruent Punchline ($CP$) (mean funniness = 1.47). The joke fragments without a counterpart--i.e., either Setup ($S$) or Punchline ($P$) alone--were considered least funny of all (the mean funniness of both was 1.22).
The dataset is entirely consistent with the view that the humour derives from incongruence due to bisociation. 

\begin{table}[ht]
  \begin{center}
  \begin{tabular}{@{} clccccrrr @{}}
    \hline \hline 
    Joke Variant & O & S & P & CS & CP & IS & IP \\ 
    \hline
    Mean Funniness & 2.70 & 1.22 & 1.22 & 1.41 & 1.47 & 2.37 & 2.12 \\
    \hline \hline
  \end{tabular}
  \end{center}
\vspace{1mm}\textbf{\refstepcounter{table}\label{tab:variants}Table \arabic{table}. }
{
The mean funniness ratings across all participants and all joke sets for each kind of joke variant. O refers to Original; S refers to Set-up only; P refers to Punchline only; CS refers to Congruent Set-up; CP refers to Congruent Punchline; IS refers to Incongruent Set-up; IP refers to Incongruent Punchline. 
}
\end{table}


\subsection{Towards a test of the QTH}

Recall that the Law of Total Probability (LTP) as represented by equation \eqref{eq:LTP} suggests that the mean funniness of a joke should be equal to the sum of its funniness as judged under all possible semantic interpretations. This is not an equality that we can directly test given our current understanding of language and how it might interplay with humour. 
However, the dataset reported here gives us some initial ways in which to consider this question. With a methodology for converting the Likert scale ratings into projective measurements of a a joke being funny or not, we can start to consider the relative frequency that an original joke is judged as funny and comparing this result with the individual components. 

\begin{landscape} 
\begin{figure}[p]
\parbox[c][\textwidth][s]{\linewidth}{%
\vfill
\centering
\includegraphics[width=22cm]{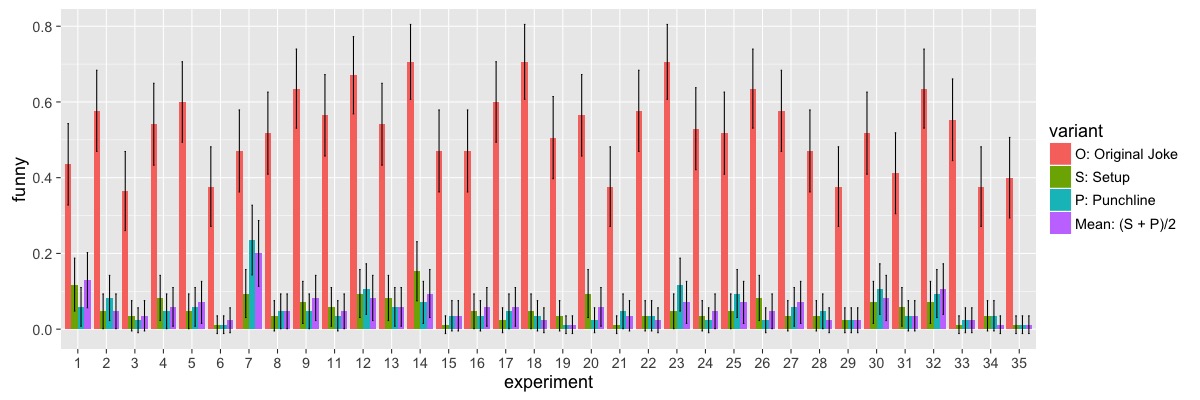}
\caption{A comparison of the frequency with which a specific joke and its fragments are considered funny for participants in the pilot trial (using a threshold value of 2.5, n=85). A mean of the set-up and the punchline variants ($V_S$ and $V_P$) is also given. Confidence intervals are set at 95\%.
}\label{fig:funninessComparison}
\vfill}
\end{figure}
\end{landscape}

We start by translating the Likert scale responses into a simplified measurement of funniness, by mapping the funniness ratings into a designation of funny or not.
In order to run a quick comparison between the relative frequencies that participants decided the  full joke ($V_O$) was funny when compared to the simple components of the joke ($V_S$ and $V_P$), we took the mean value of the components for each subject. 
Given that puns are not generally considered particularly funny (a result backed up by our participant ratings) we used a fairly low threshold value of 2.5 (i.e. if the mean was less than 2.5 then the components were judged as unfunny, and vice versa). Exploring the results of this mapping gives us the data reported in Figure~\ref{fig:funninessComparison} for the $V_O$, $V_S$ and $V_P$ variants of the jokes, listing the frequency at which participants judged the joke and subcomponents funny. A mean value for the joke fragments is also presented. All data uses confidence intervals at the 95\% level.

We see a significant discrepancy between the funniness of the original and the combined funniness of its components. This is not a terribly surprising result, jokes are not funny when the set-up is not followed by the punchline, and participants usually rated $V_S$ and $V_P$ variants as unfunny (i.e. scoring them at 1). Table~\ref{tab:long} in the Appendix shows that in the participant pool of 85, the set-up and punchline variants of the joke rarely had a mean funniness rating above 1.5. However, to extract a violation of the LTP for this scenario, we would need to construct expressions such as the following
\begin{equation}
\label{eq:ateEight}
p(F)=p(EIGHT).p(F|EIGHT)+p(ATE).p(F|ATE).
\end{equation}
How precisely could such a relationship be tested? Two forms of data are required to test whether the simple puns used in our experiment actually violate the LTP: 
\begin{enumerate}
\item {\bf Funniness ratings:} These are the probabilities regarding the probability that different components of the joke are considered funny (the whole joke ($p(F)$); just the setup ($p(F|EIGHT)$); and just the punchline ($p(F|ATE)$); and
\item {\bf Semantic probabilities:} These list the probability of EYT being interpreted as EIGHT: $p(EIGHT)$, or ATE: $p(ATE)$, within the context of the specific joke fragment.
\end{enumerate}
We have demonstrated a method for extracting the funniness ratings above. How might we obtain data for the semantic probabilities? First we must consider the precise interpretation of what these probabilities might actually be. 
Firstly, we note that it seems likely participants will interpret just a set-up or a punchline in the sense that the fragment represents. The bisociation that humour relies upon is not present for a fragment, and so a person hearing a fragment will be primed by its surrounding context towards interpreting an ambiguous word in precisely the sense intended for that fragment. Indeed, the incongruity that results from having to readjust the interpretation of the joke, and the resulting bisociation, lies at the very base of the humour that arises. Free association probabilities will not give these values. To test the LTP, it would be necessary to extract information about how a participant is interpreting core terms in the joke as it progresses; some form of nondestructive measurement is required, and a new experimental protocol will have to be defined. We reserve this for future work.

However, the significant difference between the rated funniness of the fragments and that of the original joke allows us to formulate an alternative mechanism for testing equations of the form~\eqref{eq:LTP} and \eqref{eq:ateEight}. We can do this by asking whether there is \emph{any} way in which the semantic probabilities could have values that would satisfice the LTP?
An examination of  Figure~\ref{fig:funninessComparison} for the setup and punchline variants of the jokes suggests that there is no way in which to chose semantic probabilities that will satisfy the LTP. Thus, we have preliminary evidence that humour should perhaps be treated using a quantum inspired model. 

\section{Discussion}

It would appear that there is some support for the hypothesis that the humour arising from bisociation can be modelled by a quantum inspired approach. Furthermore, the experimental results presented in section~\ref{sec:experimentsHumour} suggest that this model might more appropriate than one grounded in classical probability. However, much work remains to be completed before we can consider these findings anything but preliminary. 

Firstly, the model presented in Section~\ref{sec:quantumModel} is simple, and will need to be extended. 
 While an extension to more senses for an ambiguous element of a joke is straightforward with a move to higher dimensions, the model is currently not well suited to the set of different variants discussed in section~\ref{sec:experiment}. A model that can show how they interrelate, and how their underlying semantics affects the perceived humour in a joke is desirable. Furthermore, the funniness of the joke was simplistically represented by a projection onto the `funny'/`not funny' axis. A more theoretically grounded treatment of the Likert data is desirable. For example, the current threshold value of 2.5 was chosen somewhat arbitrarily (although could be justified by a consideration of the mean values for funniness scores reported in the Appendix --- see Table~\ref{tab:long}).  
A  more systematic way of considering the Likert scale measures to allow for a normalisation of funniness ratings at the level of an individual is also desirable. As a highly subjective phenomenon, funniness is liable to be judged by different individuals inconsistently and so it will be important that we control for this effect in comparing Likert responses among individuals. 

Considering experimental results, 
the sample size of the data set is somewhat small (85 participants), although our funniness ratings appear to be reasonably stable for this cohort. A more concerning problem revolves around the construction of a LTP relationship for our simple model. There are many alternative ways in which a LTP could be constructed for puns, and more sophisticated models need to be investigated before we can feel confident that our results do indeed demonstrate that humour \emph{must} be modelled using a quantum inspired approach. 
 In particular, we require a more sophisticated method that facilitates the extraction of data about the semantics attributed by a participant to a joke. A two stage protocol may be the answer for obtaining the necessary semantic information and so providing a more rigorously founded test of the violation of LTP. 
It would be useful to construct a systematic study of the manner in which adjusting the congruence of the set-ups and punchlines influences perception of the joke.
The quantum inspired semantic space approaches of \citet{van_rijsbergen_geometry_2004,widdows_geometry_2004} may prove fruitful in this case, as they would facilitate the creation of similarity models such as those explored by \citet{aerts_similarity_2011,pothos_structured_2015}.

In summary, humour is complex, and it will take an ongoing program of research to gradually understand the interplay between the semantics of a joke and its perceived funniness. However, at this point we might pause to consider the broader question of \emph{why} humour might be better modelled by a quantum inspired approach than by one grounded in classical probability? To this end we return to the discussion of Section~\ref{sec:quantum}. As we saw, the humour of a pun involves the bisociation of incongruent frames, i.e., re-viewing a setup frame in light of new contextual information provided by a punchline frame.
Moreover, the broader contextuality of humour means that even the funniest of jokes can become markedly unfunny if delivered in the wrong way (e.g. a monotone voice), or in the wrong situation (e.g. after receiving very bad news). Funniness is not a pre-existing `element of reality' that can be measured; it emerges from an interaction between the underlying nature of the joke, the cognitive state of the listener, and other social and environmental factors. This makes the quantum formalism an excellent candidate for modelling humour, as this interaction is well described by the concept of a vector state embedded in a space which is represented using basis states that can be reoriented according to the framing of the joke. However, this paper only provides a preliminary indication that a QTH may indeed provide a good theoretical underpinning for this complex process. Much more work remains to be done. 

\section{Conclusions}

This paper has provided a first step towards a quantum theory humour (QTH). We constructed a model where frame blends are represented in a Hilbert space spanned by two sets of basis states, one representing the ambiguous framing of a joke, and the other representing funniness. The process of `getting a joke' then consists of a dual stage scenario, where the cognitive state of a person evolves towards a re-interpretation of the meaning attributed to the joke, followed by a measurement of funniness. We conducted a study in which participants rated the funniness of jokes as well as the funniness of variants of those jokes consisting of setting or punchline by alone. The results demonstrate that the funniness of the jokes is significantly greater than that of their components, which is not particularly surprising, but does show that there is something cognitive taking place above and beyond the information content delivered in the joke. A preliminary test to see whether the humour in a joke violates the law of total probability appears to suggest that there is some reason to suppose that a quantum inspired model is indeed appropriate.

Our QTH is not proposed as an all-encompassing theory of humour; for example, it cannot explain why laughter is contagious, or why children tease each other, or why people might find it funny when someone is hit in the face with a pie (and laugh even if they know it will happen in advance). It aims to model the cognitive aspect of humour only. Moreover, despite the intuitive appeal of the approach, it is still rudimentary, and more research is needed to determine to what extent it is consistent with empirical data. Nevertheless we believe that the approach promises an exciting step toward a formal theory of humour. It is hoped that future research will build upon this modest beginning.

\section*{Acknowledgements}
This work was supported by a grant (62R06523) from the Natural Sciences and Engineering Research Council of Canada. We are grateful to Samantha Thomson who assisted with the development of the questionnaire and the collection of the data for the study reported here. 

\newpage
\bibliography{QuantumTheoryHumor}

\newpage

\section*{Appendix}
\label{sec:fullJokes}

\begin{center}
\begin{longtable}{|l|l|p{9cm}|l|}
\caption{Mean funniness ratings across all participants for the entire collection of jokes and their variants. O refers to Original; S refers to Set-up only; P refers to Punchline only; CS refers to Congruent Set-up; CP refers to Congruent Punchline; IS refers to Incongruent Set-up; IP refers to Incongruent Punchline. The CS, CP, IS, and IP data was not included in this analysis, but it is clearly consistent with the hypothesis that incongruency is important to humour.} \label{tab:long} \\
\hline 
\hline 
\multicolumn{1}{|c|}
{\textbf{Joke Set}} & \multicolumn{1}{c|}
{\textbf{Type}} & \multicolumn{1}{c|}
{\textbf{Joke}} & \multicolumn{1}{c|}
{\textbf{Funniness}} 
\\ \hline 
\endfirsthead

\multicolumn{4}{c}%
{{\bfseries \tablename\ \thetable{} -- continued from previous page.}} \\
\hline \multicolumn{1}{|c|}
{\textbf{Joke Set}} & \multicolumn{1}{c|}
{\textbf{Type}} & \multicolumn{1}{c|}
{\textbf{Joke}} & \multicolumn{1}{c|}
{\textbf{Funniness}} 
\\ \hline 
\endhead

\hline \multicolumn{4}{|r|}{{Continued on next page}} \\ \hline
\endfoot

\hline \hline
\endlastfoot

1 & O & Why was 6 afraid of 7? Because 7, 8, 9. & 2.62 \\
1 & S & Why was 6 afraid of 7? & 1.00 \\
1 & P & Because 7, 8, 9. & 1.43 \\
1 & IS & Why was the child afraid of getting older? Because 7, 8, 9. & 2.33 \\
1 & IP & Why was 6 afraid of 7? Because 7 is an odd number. & 2.90 \\
    \hline
2 & O & What do you call someone else's cheese? Nacho cheese! & 3.24 \\
    2 & S & What do you call someone else's cheese? & 1.00 \\
    2 & P & Nacho cheese! & 2.14 \\
    2 & CP & What do you call someone else's cheese? Cheese that doesn't belong to you! & 1.76 \\
    2 & IS & What did the boy say when someone tried to take his cheese? Nacho cheese! & 3.38 \\
             \hline 
	3 & O & Why did the boy bring a ladder to school? Because he wanted to go to high school! & 3.33 \\
	3 & S & Why did the boy bring a ladder to school? & 1.24 \\
	3 & P & Because he wanted to go to high school! & 1.00 \\
	3 & IS & Why did the boy smoke pot before school? Because he wanted to go to high school! & 3.48 \\
	3 & IP & Why did the boy bring the ladder to school? Because the teacher said she had high standards for him. & 2.52 \\
             \hline
	4 & O & Two jumper cables walk into a bar. The bartender says ``I'll serve you, but don't start anything!'' & 2.95\\
	4 & S & Two jumper cables walk into a bar. & 2.00 \\
	4 & P & The bartender says ``I'll serve you, but don't start anything!'' & 1.81 \\
	4 & CS & A man known for his violent tendencies walks into a bar. The bartender says ``I'll serve you, but don't start anything!'' & 1.86 \\
	4 & CP & Two jumper cables walk into a bar. Because they are being carried by a person who frequents the bar. & 1.52 \\
             \hline
    5 & O & Why is air a lot like sex? Because it is no big deal unless you're not getting any. & 3.86 \\
	5 & S & Why is air a lot like sex? & 1.71 \\
	5 & P & Because it is no big deal unless you're not getting any. & 1.24 \\
    5 & IS & Why is air a lot like food? Because it is no big deal unless you're not getting any. & 3.19 \\
    5 & IP & Why is air a lot like sex? Both are vital to the continuation of human life. & 2.76\\
            \hline
    6 & O & A guy shows up late for work. The boss yells at him, ``You should have been here at 8:30!'' He replies, ``Why? What happened at 8:30?'' & 2.52 \\
    6 & S & A guy shows up late for work. The boss yells at him, ``You should have been here at 8:30!'' & 1.00 \\
    6 & P & He replies, ``Why? What happened at 8:30?'' & 1.57\\
    6 & CS & A guy shows up late for work and his coworker says, ``You should have been here earlier, something crazy happened at 8:30!'' He replies, ``Why? What happened at 8:30?'' & 1.33\\
    6 & CS & A guy shows up late for work. The boss yells at him, ``You should have been here at 8:30!'' He replies, ``I know, I'm sorry!'' & 1.24\\
             \hline
    7 & O & You don't need a parachute to go skydiving. You need a parachute to go skydiving twice. & 3.62\\
    7 & S & You don't need a parachute to go skydiving. & 1.24\\
    7 & P & You need a parachute to go skydiving twice. & 1.81\\
    7 & CP & You don't need a parachute to go skydiving, but it helps to with the landing. & 2.57\\
    7 & IS & You need a plane to go skydiving. You need a parachute to go skydiving twice. & 2.80\\
             \hline
             
    8 & O & Want to hear a word I just made up? Plagiarism. & 3.29\\
    8 & S & Want to hear a word I just made up? & 1.14\\
    8 & P & Plagiarism. & 1.24\\
    8 & CS & Do I need to cite the word ?plagiarism'? & 2.48\\
    8 & IP & Want to hear a word I just made up? Gullible. & 3.57\\
             \hline
    9 & O & A police officer called the station and said ``I have an interesting case. A woman shot her husband for stepping on the floor she just mopped.'' ``Have you arrested this woman?'' ``No, the floor is still wet.'' & 3.62\\
    9 & S & A police officer called the station and said ``I have an interesting case. A woman shot her husband for stepping on the floor she just mopped.'' ``Have you arrested this woman?'' & 1.19\\
    9 & P & ``No, the floor is still wet.'' & 1.00\\
    9 & CS & The husband asked his wife if he could come into the kitchen after she mopped it and she said, ``No, the floor is still wet.'' & 1.19\\
    9 & CP & A police officer called the station and said ``I have an interesting case. A woman shot her husband for stepping on the floor she just mopped.'' ``Have you arrested this woman?'' ``No, she fled the scene.'' & 2.14\\
            \hline
    10 & O & Why did the cookie go to the doctor's office? Because it was feeling crummy. & 3.57\\
    10 & S & Why did the cookie go to the doctor's office? & 1.14\\
    10 & P & Because it was feeling crummy. & 1.00 \\
    10 & IS & Why did the cookie need a napkin? Because it was feeling crummy. & 2.33 \\
    10 & IP & Why did the cookie go to the doctor's office? Because it was sick. & 1.29\\
           \hline
    11 & O & What happens to a frog's car when it breaks down? It gets toad. & 2.72 \\
    11 & S & What happens to a frog's car when it breaks down? & 1.20 \\
    11 & P & It gets toad. & 1.15 \\
    11 & CP & What happens to a frog's car when it breaks down? It gets it towed. & 1.82 \\
    11 & IS & What happens when the frog parks illegally? It gets toad! &  2.94\\
    11 & IP & What does a frog say when it's car breaks down? "It croaked!" & 2.09 \\
          \hline
     12 & O & My friend thinks he is so smart. He told me an onion is the only food that makes you cry. To prove him wrong I threw a coconut at his face. & 3.20 \\
     12 & S & My friend thinks he is so smart. He told me an onion is the only food that makes you cry. & 1.35 \\
     12 & P & To prove him wrong I threw a coconut at his face. & 1.47 \\
     12 & IP & My friend thinks he is so smart. He told me an onion is the only food that makes you cry. I told him a coconut would make you cry if I threw it at him. & 2.33 \\
     12 & IS & My friend thinks he is so smart. He told me that all fruits are good for you. I threw a coconut at his face to prove him wrong. & 2.41 \\
     \hline
     13 & O & What did the duck say when she bought the lipstick? “Put it on my bill.” & 2.68 \\
13 & S & What did the duck say when she bought the lipstick? & 1.25 \\
13 & P & “Put it on my bill.” & 1.19 \\
13 & CP & What did the duck say when she bought the lipstick? “Does this colour look good on me?” & 1.25 \\
13 & CS & The bar patron asked the bartender for another drink and told him, “Put it on my bill.” & 1.22 \\
13 & IP & What did the duck say when she bought the lipstick? “Was this tested on animals?” & 2.07 \\
13 & IS & The duck asked the bartender for another drink and told him, “Put it on my bill.” & 2.69 \\
          \hline
          14 & O & In a Catholic school cafeteria, a nun places a sign in front of a pile of apples, “Only take one, God is watching”. Further down the line is a pile of cookies. A little boy makes his own sign, “Take all you want, God is watching the apples”. & 3.33 \\
14 & S & In a Catholic school cafeteria, a nun places a sign in front of a pile of apples, “Only take one, God is watching”. Further down the line is a pile of cookies. & 1.49 \\
14 & P & A little boy makes his own sign, “Take all you want, God is watching the apples”. & 1.32 \\
14 & CP & In a Catholic school cafeteria, a nun places a sign in front of a pile of apples, “Only take one, God is watching”. Further down the line is a pile of cookies with the same sign. & 1.36 \\
14 & IS & In a Catholic school cafeteria, a little boy places a sign in front of a pile of apples, “Only take one, God is watching”.  Further down the line is a pile of cookies. The boy makes another sign, “Take all you want, God is watching the apples”. & 3.05 \\
          \hline
15 & O & Can a kangaroo jump higher than the Empire State Building? Of course it can, the Empire State Building can’t jump! & 2.54 \\
15 &S & Can a kangaroo jump higher than the Empire State Building? & 1.09 \\
15 & P & Of course it can, the Empire State Building can’t jump! & 1.19 \\
15 & CP & Can a kangaroo jump higher than the Empire State Building? No, that is impossible. & 1.32 \\
15 & CS &  Which can jump higher, the Empire State Building or a kangaroo? A kangaroo, the Empire State Building can’t jump! & 2.04 \\
          \hline
16 & O & What do you call a pig that does karate? A pork chop! & 2.54 \\
16 & S & What do you call a pig that does karate? & 1.29 \\
16 & P & A pork chop! & 1.13 \\
16 & CP & What do you call a pig that does karate? A pig that does karate! & 1.52 \\
16 & CS & What is a popular cut of meat? A pork chop. & 1.45 \\
16 & IP & What do you call a pig that does karate? A kung pow piggy. & 1.87 \\
          \hline
17 & O & How do trees access the internet? They log in. & 2.84 \\ 
17 & S & How do trees access the internet? & 1.14 \\
17 & P & They log in. & 1.19 \\
17 & CP & How do trees access the internet? They don’t. & 1.78 \\
17 & CS & How does one access the internet? They log in. & 1.32 \\
          \hline
18 & O & Why can’t you trust an atom? Because they make up everything! & 3.16 \\
18 & S & Why can’t you trust an atom? & 1.21 \\
18 & P & Because they make up everything! & 1.12 \\
18 & IP & Why can’t you trust an atom? Because they’re always moving! & 1.79 \\
18 & CS & Why can’t you trust a liar? Because they make up everything! & 1.54 \\
          \hline
19 & O & What kind of nails do carpenters hate to hit? Fingernails! & 2.58 \\
19 & S & What kind of nails do carpenters hate to hit? & 1.14 \\
19 & P & Fingernails! & 1.09 \\
19 & CP & What kind of nails do carpenters hate to hit? Tough ones. & 1.58 \\
19 & CS & What kind of nails do people paint? Fingernails! & 1.40 \\
          \hline
20 & O & The energizer bunny was arrested on a charge of battery. & 2.79 \\
20 & S & The energizer bunny was arrested. & 1.33 \\
20 & P & on a charge of battery. & 1.14 \\
20 & IP & The energizer bunny was arrested on a charge of false advertisement. & 1.53 \\
20 & CS & The angry man was arrested on a charge of battery. & 1.28 \\
          \hline
21 & O & Why did the skeleton cross the road? To get to the body shop! & 2.20 \\
21 & S & Why did the skeleton cross the road? & 1.12 \\
21 & P & To get to the body shop! & 1.17 \\
21 & IP & Why did the skeleton cross the road? To stretch his legs. &  1.41\\
21 & IS & Why did the invisible man cross the road? To get to the body shop! & 2.13 \\
          \hline
22 & O & Why did the fish blush? Because it saw the ocean’s bottom! & 2.82 \\
22 & S & Why did the fish blush? & 1.16 \\
22 & P & Because it saw the ocean’s bottom! & 1.14 \\
22 & IP & Why did the fish blush? Because it puffed in front of his friends. & 1.69 \\
22 & IS & Why did the surfer crash? Because he saw the ocean’s bottom! & 2.12 \\
          \hline
23 & O & What do you call a fake noodle? An impasta! & 3.01 \\
23 & S & What do you call a fake noodle? & 1.24 \\
23 & P & An impasta! & 1.42 \\
23 & IP & What do you call a fake noodle? A lack-aroni. & 2.15 \\
23 & IS & What did the mobster say when he found out his friend was a traitor? An impasta! & 1.81 \\
          \hline
24 & O & What do you call an alligator in a vest? An investigator! & 2.73 \\
24 & S & What do you call an alligator in a vest? & 1.19 \\
24 & P & An investigator! & 1.13 \\
24 & IP & What do you call an alligator in a vest? A well-dressed reptile. & 1.86 \\
24 & IS & What do you call an alligator that is a detective? An investigator! & 2.54 \\
          \hline
25 & O & What do you call a pile of kittens? A meowntain! & 2.58 \\
25 & S & What do you call a pile of kittens? & 1.16 \\
25 & P & A meowntain! & 1.34 \\
25 & IP & What do you call a pile of kittens? Mount Kili-meow-jaro. & 2.61 \\
25 & IS & What do you call a mountain for cats? A meowtain! &2.51 \\
          \hline
26 & O & What do you get from a pampered cow? Spoiled milk! & 2.80 \\
26 & S & What do you get from a pampered cow? & 1.24 \\
26 & P & Spoiled milk! & 1.16 \\
26 & IP & What do you get from a pampered cow? Udder sass! & 2.72 \\
26 & CS & What do you get when you leave out milk overnight? Spoiled milk! & 1.51 \\
          \hline
27 & O & How do you make holy water? Boil the hell out of it! & 2.96 \\
27 & S & How do you make holy water? & 1.18 \\
27 & P & Boil the hell out of it! & 1.24 \\
27 & IP & How do you make holy water? Hole punch it! & 1.85 \\
27 & CS & How do you make water really hot? Boil the hell out of it! & 1.62 \\
27 & CP & How do you make holy water? Get a priest to bless it. & 1.41 \\
          \hline 
28 & O & How do you make an octopus laugh? With ten-tickles! & 2.58 \\
28 & S & How do you make an octopus laugh? & 1.15 \\
28 & P & With ten-tickles! & 1.18 \\
28 & CP & How do you make an octopus laugh? Tell it jokes. & 1.50 \\
28 & IS & How many tickles does it take to make an octopus laugh? Ten-tickles! & 2.78 \\
          \hline
29 & O & What do you call a boy who finally stood up to the bullies? An ambulance. & 2.07 \\
29 & S & What do you call a boy who finally stood up to the bullies? & 1.08 \\
29 & P & An ambulance. & 1.06 \\
29 & CP & What do you call a boy who finally stood up to the bullies? A brave person. & 1.26 \\
29 & CS & What is the name of an emergency vehicle? An ambulance. & 1.18 \\
29 & IP & What do you call a boy who finally stood up to the bullies? An idiot! & 1.69 \\
          \hline
30 & O & Did you hear the one about the geologist? He took his wife for granite so she left him! & 2.67 \\
30 & S & Did you hear the one about the geologist? & 1.22 \\
30 & P & He took his wife for granite so she left him! & 1.38 \\
30 & IP & Did you hear the one about the geologist? He felt that no one appreciated his work; they took him for granite! & 2.53 \\
30 & IS & What happened to The Rock’s marriage? He took his wife for granite so she left him! & 2.76 \\
          \hline
31 & O & What did the tailor think of her new job? It was sew-sew. & 2.45 \\
31 & S & What did the tailor think of her new job? & 1.18 \\
31 & P & It was sew-sew. & 1.15 \\
31 & IP & What did the tailor think of her new job? She loved her new coworkers, they left her in stitches! & 2.21 \\
31 & IS & What did the sewing teacher say about her student’s skills? It was sew-sew. & 2.31 \\
          \hline
32 & O & I tried to catch some fog earlier. I mist. & 2.85 \\
32 & S & I tried to catch some fog earlier. & 1.26 \\
32 & P & I mist. & 1.36 \\
32 & CP & I tried to catch some fog earlier. It was hard! & 1.48 \\
32 & IS & I couldn’t see where I was throwing my ball through the fog so I mist. & 2.48 \\
          \hline
33 & O & What happened to the plant in math class? It grew square roots! & 2.73 \\
33 & S & What happened to the plant in math class? & 1.10 \\
33 & P & It grew square roots! & 1.14 \\
33 & IP & What happened to the plant in math class? It died of boredom! & 1.93 \\
33 & IS & What happened when the tree went to college? It grew square roots! & 2.34 \\
          \hline
34 & O & What kind of bus can you never enter? A syllabus. & 2.29 \\
34 & S & What kind of bus can you never enter? & 1.11 \\
34 & P & A syllabus. & 1.10 \\
34 & IP & What kind of bus can you never enter? A rhombus! & 2.20\\
34 & CS & What do you look at to learn about a class? A syllabus. & 1.34 \\
          \hline
35 & O & What kind of dress can’t be worn? An address! & 2.43 \\
35 & S & What kind of dress can’t be worn? & 1.09 \\
35 & P & An address! & 1.10 \\
35 & CP & What kind of dress can’t be worn? One that is too small. & 1.53 \\
35 & IS & Where can you find a dressmaker? Their address! & 1.91 \\
            \hline
    

\end{longtable}
\end{center}

\end{document}